\documentclass[longauth]{aa}  

\usepackage{stfloats}

\usepackage{graphicx}   
\usepackage[fleqn]{amsmath}
\usepackage{array, amssymb}	
\usepackage[utf8]{inputenc} 
\usepackage[T1]{fontenc}    
\usepackage{url}            
\usepackage{booktabs}       
\usepackage{lmodern}
\usepackage{multirow}
\usepackage[colorlinks,allcolors=blue]{hyperref}
\usepackage{orcidlink} 

\usepackage{xspace}
\newcommand{\gray}{$\gamma$-ray\xspace}
\newcommand{\grays}{$\gamma$-rays\xspace}
\newcommand{\RXJ}{RX\,J0852.0$-$4622\xspace}
\newcommand{\hessj}{HESS\,J0852$-$463\xspace}
\newcommand{\psr}{PSR\,J0855$-$4644\xspace}
\newcommand{\hess}{H.E.S.S.\xspace}

\begin{document} 

  \title{H.E.S.S. detection of the \psr nebula}

  \author{
{\small
F.~Aharonian\inst{\ref{YSU},\ref{MPIK},\ref{DIAS}}\orcidlink{0000-0003-1157-3915} 
\and H.~Ashkar\inst{\ref{LLR}}\orcidlink{0000-0002-2153-1818} 
\and M.~Backes\inst{\ref{UNAM},\ref{NWU}}\orcidlink{0000-0002-9326-6400} 
\and R.~Batzofin\inst{\ref{UP}}\orcidlink{0000-0002-5797-3386} 
\and Y.~Becherini\inst{\ref{APC}}\orcidlink{0000-0002-2115-2930} 
\and D.~Berge\inst{\ref{DESY},\ref{HUB}}\orcidlink{0000-0002-2918-1824} 
\and K.~Bernl\"ohr\inst{\ref{MPIK}}\orcidlink{0000-0001-8065-3252} 
\and M.~B\"ottcher\inst{\ref{NWU}}\orcidlink{0000-0002-8434-5692} 
\and C.~Boisson\inst{\ref{LUX}}\orcidlink{0000-0001-5893-1797} 
\and J.~Bolmont\inst{\ref{LPNHE}}\orcidlink{0000-0003-4739-8389} 
\and F.~Brun\inst{\ref{IRFU}}\orcidlink{0000-0003-0770-9007} 
\and B.~Bruno\inst{\ref{ECAP}}\orcidlink{0000-0002-0792-6311} 
\and C.~Burger-Scheidlin\inst{\ref{DIAS}}\orcidlink{0000-0002-7239-2248} 
\and T.~Bylund\inst{\ref{LUX}}\orcidlink{0000-0003-2946-1313} 
\and S.~Casanova\inst{\ref{IFJPAN}}\orcidlink{0000-0002-6144-9122} 
\and D.~Cecchin~Momesso\inst{\ref{ECAP}}\orcidlink{0000-0001-6709-509X} 
\and J.~Celic\inst{\ref{ECAP}}
\and M.~Cerruti\inst{\ref{APC}}\orcidlink{0000-0001-7891-699X} 
\and A.~Chen\inst{\ref{Wits}}\orcidlink{0000-0001-6425-5692} 
\and M.~Chernyakova\inst{\ref{DCU},\ref{DIAS}}\orcidlink{0000-0002-9735-3608} 
\and J. O.~Chibueze\inst{\ref{NWU},\ref{UNAM}}\orcidlink{0000-0002-9875-7436} 
\and O.~Chibueze\inst{\ref{NWU}}\orcidlink{0000-0001-8601-2675} 
\and B.~Cornejo\inst{\ref{IRFU}}\orcidlink{0009-0003-0039-0483} 
\and G.~Cotter\inst{\ref{UOX}}\orcidlink{0000-0002-9975-1829} 
\and G.~Cozzolongo\inst{\ref{ECAP}}
\and J.~de~Assis~Scarpin\inst{\ref{LLR}}\orcidlink{0009-0004-4411-236X} 
\and M.~de~Bony~de~Lavergne\inst{\ref{IRFU},\ref{CPPM}}\orcidlink{0000-0002-4650-1666} 
\and M.~de~Naurois\inst{\ref{LLR}}\orcidlink{0000-0002-7245-201X} 
\and E.~de~O\~na~Wilhelmi\inst{\ref{DESY}}\orcidlink{0000-0002-5401-0744} 
\and A.~G.~Delgado~Giler\inst{\ref{HUB}}\orcidlink{0000-0003-2190-9857} 
\and J.~Devin\inst{\ref{LUPM}}\orcidlink{0000-0003-1018-7246} 
\and A.~Dmytriiev\inst{\ref{NWU}}\orcidlink{0000-0003-0102-5579} 
\and K.~Egberts\inst{\ref{UP}}\orcidlink{0009-0000-5511-7060} 
\and K.~Egg\inst{\ref{ECAP}}\orcidlink{0009-0002-4238-034X} 
\and J.-P.~Ernenwein\inst{\ref{CPPM}}
\and C.~Esca\~{n}uela~Nieves\inst{\ref{MPIK}}\orcidlink{0000-0002-7297-8126} 
\and P.~Fauverge\inst{\ref{LP2I}}\orcidlink{0009-0006-1613-6633} 
\and K.~Feijen\inst{\ref{APC}}{\large$^{,\star}$}\orcidlink{0000-0003-1476-3714} 
\and M.~D.~Filipovic\inst{\ref{Sydney}}\orcidlink{0000-0002-4990-9288} 
\and G.~Fontaine\inst{\ref{LLR}}\orcidlink{0000-0002-6443-5025} 
\and S.~Funk\inst{\ref{ECAP}}\orcidlink{0000-0002-2012-0080} 
\and S.~Gabici\inst{\ref{APC}}
\and Y.A.~Gallant\inst{\ref{LUPM}}
\and M.~Genaro\inst{\ref{ECAP}}\orcidlink{0000-0003-3461-1929} 
\and J.F.~Glicenstein\inst{\ref{IRFU}}\orcidlink{0000-0003-2581-1742} 
\and J.~Glombitza\inst{\ref{ECAP}}\orcidlink{0000-0001-9683-4568} 
\and P.~Goswami\inst{\ref{LSW}}\orcidlink{0000-0001-5430-4374} 
\and M.-H.~Grondin\inst{\ref{LP2I}}\orcidlink{0000-0002-8383-251X} 
\and L.~Heckmann\inst{\ref{APC}}\orcidlink{0000-0002-6653-8407} 
\and B.~Heß\inst{\ref{IAAT}}\orcidlink{0009-0004-9999-171X} 
\and W.~Hofmann\inst{\ref{MPIK}}\orcidlink{0000-0001-8295-0648} 
\and T.~L.~Holch\inst{\ref{DESY}}\orcidlink{0000-0001-5161-1168} 
\and M.~Holler\inst{\ref{Innsbruck}}\orcidlink{0000-0002-0107-8657} 
\and D.~Horns\inst{\ref{UHAM}}\orcidlink{0000-0003-1945-0119} 
\and M.~Jamrozy\inst{\ref{OAUJ}}\orcidlink{0000-0002-0870-7778} 
\and F.~Jankowsky\inst{\ref{LSW}}
\and A.~Jardin-Blicq\inst{\ref{LP2I}}\orcidlink{0000-0002-6738-9351} 
\and I.~Jaroschewski\inst{\ref{IRFU}}\orcidlink{0000-0001-5180-2845} 
\and D.~Jimeno\inst{\ref{DESY}}\orcidlink{0009-0001-2499-9467} 
\and I.~Jung-Richardt\inst{\ref{ECAP}}
\and E.~Kasai\inst{\ref{UNAM}}\orcidlink{0000-0001-9696-7221} 
\and K.~Katarzy{\'n}ski\inst{\ref{NCUT}}\orcidlink{0000-0002-8806-4863} 
\and D.~Kerszberg\inst{\ref{LPNHE}}\orcidlink{0000-0002-5289-1509} 
\and B. Khélifi\inst{\ref{APC}}{\large$^{,\star}$}\orcidlink{0000-0001-6876-5577} 
\and W.~Klu{\'z}niak\inst{\ref{NCAC}}
\and N.~Komin\inst{\ref{LUPM},\ref{Wits}}\orcidlink{0000-0003-3280-0582} 
\and K.~Kosack\inst{\ref{IRFU}}{\large$^{,\star}$}\orcidlink{0000-0001-8424-3621} 
\and D.~Kostunin\inst{\ref{DESY}}\orcidlink{0000-0002-0487-0076} 
\and R.G.~Lang\inst{\ref{ECAP}}\orcidlink{0000-0003-0492-5628} 
\and S.~Lazarevi\'c\inst{\ref{Sydney}}\orcidlink{0000-0001-6109-8548} 
\and A.~Lemi\`ere\inst{\ref{APC}}\orcidlink{0000-0002-6682-7188} 
\and M.~Lemoine-Goumard\inst{\ref{LP2I}}\orcidlink{0000-0002-4462-3686} 
\and J.-P.~Lenain\inst{\ref{LPNHE}}\orcidlink{0000-0001-7284-9220} 
\and P.~Liniewicz\inst{\ref{OAUJ}}\orcidlink{0009-0008-3575-3965} 
\and A.~Luashvili\inst{\ref{NWU}}\orcidlink{0000-0003-4384-1638} 
\and J.~Mackey\inst{\ref{DIAS}}\orcidlink{0000-0002-5449-6131} 
\and D.~Malyshev\inst{\ref{IAAT}}\orcidlink{0000-0001-9689-2194} 
\and D.~Malyshev\inst{\ref{ECAP}}\orcidlink{0000-0002-9102-4854} 
\and V.~Marandon\inst{\ref{IRFU}}\orcidlink{0000-0001-9077-4058} 
\and M.~G.~F.~Mayer\inst{\ref{ECAP}}\orcidlink{0000-0002-9771-9841} 
\and A.~Mehta\inst{\ref{DESY}}
\and A.M.W.~Mitchell\inst{\ref{ECAP}}\orcidlink{0000-0003-3631-5648} 
\and R.~Moderski\inst{\ref{NCAC}}\orcidlink{0000-0002-8663-3882} 
\and L.~Mohrmann\inst{\ref{MPIK}}\orcidlink{0000-0002-9667-8654} 
\and E.~Moulin\inst{\ref{IRFU}}\orcidlink{0000-0003-4007-0145} 
\and J.~Niemiec\inst{\ref{IFJPAN}}\orcidlink{0000-0001-6036-8569} 
\and P.~O'Brien\inst{\ref{Leicester}}\orcidlink{0000-0002-5128-1899} 
\and L.~Olivera-Nieto\inst{\ref{GRAPPA}}\orcidlink{0000-0002-9105-0518} 
\and M.O.~Moghadam\inst{\ref{UP}}\orcidlink{0009-0003-2479-1863} 
\and S.~Panny\inst{\ref{Innsbruck}}\orcidlink{0000-0001-5770-3805} 
\and M.~Panter\inst{\ref{MPIK}}
\and R.D.~Parsons\inst{\ref{HUB}}\orcidlink{0000-0003-3457-9308} 
\and P.~Pichard\inst{\ref{APC}}\orcidlink{0009-0005-9803-0762} 
\and T.~Preis\inst{\ref{Innsbruck}}\orcidlink{0009-0001-7110-6764} 
\and G.~P\"uhlhofer\inst{\ref{IAAT}}\orcidlink{0000-0003-4632-4644} 
\and M.~Punch\inst{\ref{APC}}\orcidlink{0000-0002-4710-2165} 
\and A.~Quirrenbach\inst{\ref{LSW}}
\and A.~Reimer\inst{\ref{Innsbruck}}\orcidlink{0000-0001-8604-7077} 
\and O.~Reimer\inst{\ref{Innsbruck}}\orcidlink{0000-0001-6953-1385} 
\and I.~Reis\inst{\ref{IRFU}}\orcidlink{0000-0002-0771-3332} 
\and Q.~Remy\inst{\ref{MPIK}}\orcidlink{0000-0002-8815-6530} 
\and H.~X.~Ren\inst{\ref{MPIK}}\orcidlink{0000-0003-0221-2560} 
\and B.~Reville\inst{\ref{MPIK}}\orcidlink{0000-0002-3778-1432} 
\and F.~Rieger\inst{\ref{MPIK}}\orcidlink{0000-0003-1334-2993} 
\and G.~Rowell\inst{\ref{Adelaide}}\orcidlink{0000-0002-9516-1581} 
\and B.~Rudak\inst{\ref{NCAC}}\orcidlink{0000-0003-0452-3805} 
\and K.~Sabri\inst{\ref{LUPM}}
\and V.~Sahakian\inst{\ref{YPI}}\orcidlink{0000-0003-1198-0043} 
\and M.~Sasaki\inst{\ref{ECAP}}\orcidlink{0000-0001-5302-1866} 
\and F.~Sch\"ussler\inst{\ref{IRFU}}\orcidlink{0000-0003-1500-6571} 
\and J.N.S.~Shapopi\inst{\ref{UNAM}}\orcidlink{0000-0002-7130-9270} 
\and W.~Si~Said\inst{\ref{LLR}}\orcidlink{0009-0007-6555-6893} 
\and {\L.}~Stawarz\inst{\ref{OAUJ}}\orcidlink{0000-0002-7263-7540} 
\and R.~Steenkamp\inst{\ref{UNAM}}\orcidlink{0009-0009-4130-977X} 
\and S.~Steinmassl\inst{\ref{MPIK}}\orcidlink{0000-0002-2865-8563} 
\and T.~Tanaka\inst{\ref{Konan}}\orcidlink{0000-0002-4383-0368} 
\and A.M.~Taylor\inst{\ref{DESY}}\orcidlink{0000-0001-9473-4758} 
\and G.~L.~Taylor\inst{\ref{LSW}}\orcidlink{0009-0001-8062-036X} 
\and R.~Terrier\inst{\ref{APC}}{\large$^{,\star}$}\orcidlink{0000-0002-8219-4667} 
\and Y.~Tian\inst{\ref{DESY}}\orcidlink{0009-0005-7165-3791} 
\and M.~Tsirou\inst{\ref{DESY}}\orcidlink{0000-0003-3417-1425} 
\and T.~Unbehaun\inst{\ref{MPIK}}\orcidlink{0000-0002-7378-4024} 
\and C.~van~Eldik\inst{\ref{ECAP}}\orcidlink{0000-0001-9669-645X} 
\and M.~Vecchi\inst{\ref{Groningen}}\orcidlink{0000-0002-5338-6029} 
\and C.~Venter\inst{\ref{NWU}}\orcidlink{0000-0002-2666-4812} 
\and J.~Vink\inst{\ref{GRAPPA}}\orcidlink{0000-0002-4708-4219} 
\and V.~Voitsekhovskyi\inst{\ref{GRAPPA}}\orcidlink{0000-0002-3906-4840} 
\and T.~Wach\inst{\ref{MPIK}}\orcidlink{0009-0008-4658-7405} 
\and S.J.~Wagner\inst{\ref{LSW}}\orcidlink{0000-0002-7474-6062} 
\and A.~Wierzcholska\inst{\ref{IFJPAN},\ref{LSW}}\orcidlink{0000-0003-4472-7204} 
\and M.~Zacharias\inst{\ref{LSW},\ref{NWU}}\orcidlink{0000-0001-5801-3945} 
\and A.~Zech\inst{\ref{LUX}}
\and W.~Zhong\inst{\ref{DESY}}\orcidlink{0000-0003-3717-2861} 
\\ (H.E.S.S.\ Collaboration)
\and \\ F.~Acero \inst{\ref{CEA},\ref{IAC}}\orcidlink{0000-0002-6606-2816} \and L.~Giunti \inst{\ref{APC},\ref{CMCC}}\orcidlink{0000-0002-3395-3647}
}
}

\institute{
Yerevan State University, 1 Alek Manukyan St, Yerevan 0025, Armenia \label{YSU}
\and Max-Planck-Institut für Kernphysik, P.O. Box 103980, D 69029 Heidelberg, Germany \label{MPIK}
\and Astronomy \& Astrophysics Section, School of Cosmic Physics, Dublin Institute for Advanced Studies, DIAS Dunsink Observatory, Dublin D15 XR2R, Ireland \label{DIAS}
\and Laboratoire Leprince-Ringuet, École Polytechnique, CNRS, Institut Polytechnique de Paris, F-91128 Palaiseau, France \label{LLR}
\and University of Namibia, Department of Physics, Private Bag 13301, Windhoek 10005, Namibia \label{UNAM}
\and Centre for Space Research, North-West University, Potchefstroom 2520, South Africa \label{NWU}
\and Institut für Physik und Astronomie, Universität Potsdam, Karl-Liebknecht-Strasse 24/25, D 14476 Potsdam, Germany \label{UP}
\and Université Paris Cité, CNRS, Astroparticule et Cosmologie, F-75013 Paris, France \label{APC}
\and Deutsches Elektronen-Synchrotron DESY, Platanenallee 6, 15738 Zeuthen, Germany \label{DESY}
\and Institut für Physik, Humboldt-Universität zu Berlin, Newtonstr. 15, D 12489 Berlin, Germany \label{HUB}
\and LUX, Observatoire de Paris, Université PSL, CNRS, Sorbonne Université, 5 Pl. Jules Janssen, 92190 Meudon, France \label{LUX}
\and Sorbonne Université, CNRS/IN2P3, Laboratoire de Physique Nucléaire, et de Hautes Energies, LPNHE, 4 place Jussieu, 75005 Paris, France \label{LPNHE}
\and IRFU, CEA, Université Paris-Saclay, F-91191 Gif-sur-Yvette, France \label{IRFU}
\and Friedrich-Alexander-Universität Erlangen-Nürnberg, Erlangen Centre for Astroparticle Physics, Nikolaus-Fiebiger-Str. 2, 91058 Erlangen, Germany \label{ECAP}
\and Instytut Fizyki Jac{a}drowej PAN, ul. Radzikowskiego 152, ul. Radzikowskiego 152, 31-342 Kraków, Poland \label{IFJPAN}
\and School of Physics, University of the Witwatersrand, 1 Jan Smuts Avenue, Braamfontein, Johannesburg, 2050, South Africa \label{Wits}
\and School of Physical Sciences and Centre for Astrophysics \& Relativity, Dublin City University, Glasnevin, Dublin D09 W6Y4, Ireland \label{DCU}
\and University of Oxford, Department of Physics, Denys Wilkinson Building, Keble Road, Oxford OX1 3RH, UK, United Kingdom \label{UOX}
\and Aix Marseille Université, CNRS/IN2P3, CPPM, Marseille, France \label{CPPM}
\and Laboratoire Univers et Particules de Montpellier, Université Montpellier, CNRS/IN2P3, CC 72, Place Eugène Bataillon, F-34095 Montpellier Cedex 5, France \label{LUPM}
\and Université Bordeaux, CNRS, LP2I Bordeaux, UMR 5797, F-33170 Gradignan, France \label{LP2I}
\and School of Science, Western Sydney University, Locked Bag 1797, Penrith South DC, NSW 2751, Australia \label{Sydney}
\and Landessternwarte, Universität Heidelberg, Königstuhl, D 69117 Heidelberg, Germany \label{LSW}
\and Institut für Astronomie und Astrophysik, Universität Tübingen, Sand 1, D 72076 Tübingen, Germany \label{IAAT}
\and Universität Innsbruck, Institut für Astro- und Teilchenphysik, Technikerstraße 25, 6020 Innsbruck, Austria \label{Innsbruck}
\and Universität Hamburg, Institut für Experimentalphysik, Luruper Chaussee 149, D 22761 Hamburg, Germany \label{UHAM}
\and Obserwatorium Astronomiczne, Uniwersytet Jagielloński, ul. Orla 171, 30-244 Kraków, Poland \label{OAUJ}
\and Institute of Astronomy, Faculty of Physics, Astronomy and Informatics, Nicolaus Copernicus University, Grudziadzka 5, 87-100 Torun, Poland \label{NCUT}
\and Nicolaus Copernicus Astronomical Center, Polish Academy of Sciences, ul. Bartycka 18, 00-716 Warsaw, Poland \label{NCAC}
\and University of Leicester, School of Physics and Astronomy, University Road, Leicester, LE1 7RH, United Kingdom \label{Leicester}
\and GRAPPA, Anton Pannekoek Institute for Astronomy, University of Amsterdam, Science Park 904, 1098 XH Amsterdam, The Netherlands \label{GRAPPA}
\and School of Physical Sciences, University of Adelaide, Adelaide 5005, Australia \label{Adelaide}
\and Yerevan Physics Institute, 2 Alikhanian Brothers St., 0036 Yerevan, Armenia \label{YPI}
\and Department of Physics, Konan University, 8-9-1 Okamoto, Higashinada, Kobe, Hyogo 658-8501, Japan \label{Konan}
\and Kapteyn Astronomical Institute, University of Groningen, Landleven 12, 9747 AD Groningen, The Netherlands \label{Groningen}
\and Universit\'e Paris-Saclay, Universit\'e Paris Cit\'e, CEA, CNRS, AIM, 91191 Gif-sur-Yvette, France \label{CEA}
\and  FSLAC IRL 2009, CNRS/IAC, La Laguna, Tenerife, Spain \label{IAC}
\and Now at: Euro-Mediterranean Center on Climate Change (CMCC), Lecce, Italy \label{CMCC}
}

  \date{Received 1 December 2025}
  \offprints{
  \protect\\\email{\href{mailto:contact.hess@hess-experiment.eu}{contact.hess@hess-experiment.eu};}
  \protect\\\protect $^{\star}$ Corresponding authors}

  \abstract
   {\hessj is a TeV \gray source located in the Galactic plane. The region consists of a supernova remnant (SNR, \RXJ) with a shell-like morphology, commonly referred to as Vela Junior, and a pulsar named \psr. Pulsars are among the most efficient leptonic accelerators in our Galaxy, making this region particularly interesting to study.}
   {We utilise the most recent data taken by the High Energy Stereoscopic System (\hess) to investigate any \gray emission associated with the pulsar in this region, \psr.}
   {We applied a full forward-folding method on the \hess data. Utilising 3D modelling techniques, we evaluated the TeV \gray emission towards the various components of this complex system.} 
   {The distinct energy-dependent morphology observed in our data motivates further investigation of this source. We resolved the emission in the Vela Junior region into various components, several of which correspond to the SNR itself. In particular, we find a new extended component coincident with the position of \psr. The spectrum follows a power-law distribution with a best-fit index of $\Gamma_E = 1.81\pm0.07_\textrm{stat}$, which differs from the properties of the surrounding \gray emission of the Vela Junior SNR. A one-zone leptonic joint fit between the X-rays (from XMM-Newton) and \grays (from \hess) leads to a lower limit on the magnetic field of $1.6\mu$G and a spectral index of $\alpha=1.88\pm0.01$, in line with expectations of pulsar wind nebulae (PWNe).}
   {In this paper, we report the first detection of the PWN of \psr at TeV energies with the \hess experiment, at a significance of $12.2\sigma$. This is attributed to the advanced techniques of the 3D analysis. Based on the pulsar’s characteristics, its PWN is consistent with the known TeV PWNe population in the Galaxy.}
   

   \authorrunning{H.E.S.S. Collaboration}

   \maketitle

\section{Introduction}

In the centre of the Vela constellation, close to the south-eastern edge of the Vela supernova remnant (SNR), lies another Galactic SNR designated \RXJ (G266.2$-$1.2, but most often referred to as Vela Junior). Its age is estimated to be between 1700\,yr and 4300\,yr \citep{2008_Katsuda}.
This SNR displays a clear shell-type morphology, first revealed in X-rays \citep{1998_Aschenbach} and later detected at both \gray \citep{2011_tanaka_fermi, 2007_hess_snr} and radio \citep{2018_Maxted_radio} wavelengths. 

In 2018, a deep observation campaign was performed by the High Energy Stereoscopic System (\hess) at TeV energies, resolving a series of small-scale morphological substructures along the SNR shell \citep{2018_hess_deep}. 
One such substructure, located on the south-eastern rim of the SNR, is positionally coincident with an energetic (\mbox{$\dot{E}=1.1\times10^{36}$ erg s$^{-1}$}) pulsar named \psr \citep{2003_psrs}, with a characteristic age of $\tau_c=140\,$kyr. 
The pulsar is unlikely to be physically connected to Vela Junior and thus not to have originated in the same supernova event \citep{2013_acero_xrays}.
This leads to the hypothesis that some of the TeV emission typically attributed to Vela Junior could actually originate from a hypothetical \gray bright pulsar wind nebula (PWN) surrounding \psr. This currently unresolved issue motivated the work presented in this paper. 

\psr is characterised by remarkable multi-wavelength features: in X-rays it has a compact ($\sim10$'') core with asymmetric double-jet morphology \citep{2017_Maitra_xrays} surrounded by a 150'' size diffuse PWN \citep[as determined by dedicated XMM-Newton observations,][]{2013_acero_xrays}. 
The spectral index in the X-ray band is significantly steeper towards the outer region of the nebula ($\Gamma=2.01{\pm0.04}$) than the inner region ($\Gamma=1.70{^{+0.07}_{-0.06}}$). In the radio band, \psr exhibits a partial ring-like structure and two faint tail-like features indicative of a supersonic bow shock nebula \citep{2018_Maitra_radio}. The authors conclude that this is a radio PWN of \psr. 
The observed features are likely due to synchrotron cooling of relativistic $e^\pm$ pairs, which are efficiently accelerated both in the pulsar's wind and at the termination shock between the wind and the external diffuse nebula \citep[see][for a review]{Gaensler_2006}. 
The same electron population can also produce \gray emission through inverse-Compton (IC) up-scattering. Pulsar wind nebulae are a dominant source of TeV \grays \citep{hgps_2018}, in some cases having been detected up to hundreds of TeV \citep{lhaaso}, indicating that pulsars are among the most efficient leptonic accelerators in our Galaxy.

Based on its high spin-down luminosity and its proximity to the Solar System \citep[$d\le\,900$\,pc,][]{2013_acero_xrays}, \psr is likely to emit TeV \gray emission as a PWN at a level detectable by the current generation of ground-based \gray observatories. This object has never before been identified at such high energies, largely due to its proximity to the bright Vela Junior shell and the limitations of previous analysis techniques. 

Previous \hess observations suggested a possible hardening of the \gray spectrum coinciding with the position of \psr. However, the statistical uncertainties resulting from the limited on-source exposure time and the analysis techniques employed prevented the authors from confirming significant deviations in any particular region of the Vela Junior shell \citep{2018_hess_deep}.

In this work, we report the detection of the PWN of \psr at TeV energies with the \hess experiment. 
Compared to previous publications, we use additional data, including 55 hours taken with the upgraded \hess camera system \citep{camera_upgrade_2020} and employ improved analysis techniques \citep{gammapy_2023}. 
These advancements allow us to distinguish the PWN of \psr from the bright Vela Junior shell for the first time, based on its significantly different morphology and \gray spectrum.

\section{H.E.S.S. observations and data analysis}
\label{HESS}
This paper makes use of data collected between January 2004 and February 2021 with the initial configuration of the \hess experiment, an array of four 12$\,$m diameter Imaging Atmospheric Cherenkov Telescopes (IACTs) located in the Khomas Highlands of Namibia. An additional 28$\,$m diameter IACT was added in 2012; however, we did not utilise its observational data in this work.
The data were processed with the \hess analysis package (HAP) to produce lists of $\gamma$-like events and dedicated instrument response functions (IRFs). The Hillas shower reconstruction technique \citep{hillas} and the multi-variate analysis (MVA) for $\gamma$/hadron separation \citep{ohm_2009,becherini} were adopted, with standard cuts optimised for Galactic source analysis \citep{2015_Khelifi_hapfr}. 
The event lists and IRFs were exported to FITS files complying with the common data format developed by \cite{gadf_deil} and used to perform a high-level analysis with Gammapy~v1.3 \citep{gammapy_2023, gammapy_zenodo_13}. A complete analysis cross-check was also performed, using an alternative low-level chain of calibration, reconstruction, and $\gamma$-hadron separation methods \citep{parsons_2014, 2019_lars_methods}, leading to consistent results within the typical H.E.S.S. systematic uncertainties as per \cite{systematics_2006}.
All standard-quality observations centred within $3^{\circ}$ of the nominal centre of Vela Junior with a maximum zenith angle of $50^{\circ}$were selected to cover the \hess field of view (FoV) (radius of $\sim2.5^{\circ}$ at 1\,TeV), resulting in a total observation time of $\sim130\,$hours on-source after cuts. 

As opposed to the standard morphological and spectral analyses presented in \citet{2018_hess_deep}, we performed a 3D binned likelihood analysis (hereafter 3D analysis), consisting of a simultaneous spectro-morphological modelling of the total \gray (and $\gamma$-like cosmic-ray background) emission in the source region \citep{2019_lars_methods, 2021_Giunti_hess}. 
To limit analysis systematics, observations obtained before and after the 2016 camera upgrade \citep{camera_upgrade_2020} were not stacked together. Instead, they were analysed simultaneously in a joint-likelihood approach, assuming the same source model for the two datasets but different IRFs. 
To fully enclose the complex source emission and guarantee robust constraints on the background normalisation, we adopted a $6^{\circ}$ by $6^{\circ}$ analysis region centred on Vela Junior, divided into $0.02^{\circ}$ squared pixels. The axis in reconstructed energy was divided into 13 logarithmically spaced bins between 0.7 and 100\,TeV. 
The energy threshold of 0.7\,TeV was derived based on the analysis configuration and to account for mis-modelling of the background template due to atmospheric and instrumental variations.
The cosmic ray background level in the region was evaluated from a 3D FoV background model adapted to the specific conditions of each observation \citep{2019_lars_methods, 2021_Giunti_hess}.

Following \citet{2021_Giunti_hess}, we started from a background-only hypothesis and iteratively injected parametric model components until further additions yielded no statistically significant improvement in the regional description. For each component, we tested different spatial shapes (e.g. circular vs elliptical Gaussian) and spectral functions (e.g. power law with and without a high-energy cutoff). 
We additionally performed a morphological analysis of the \gray emission in the region, integrating the signal in different energy bands. This approach allowed us to confirm the conclusions obtained from the 3D analysis by verifying the significant energy-dependent morphology of the \gray flux across the Vela Junior shell.

\section{Results}
\label{results}

\subsection{Energy resolved flux maps}
\begin{figure*}
\centering
\includegraphics[width=2\columnwidth]{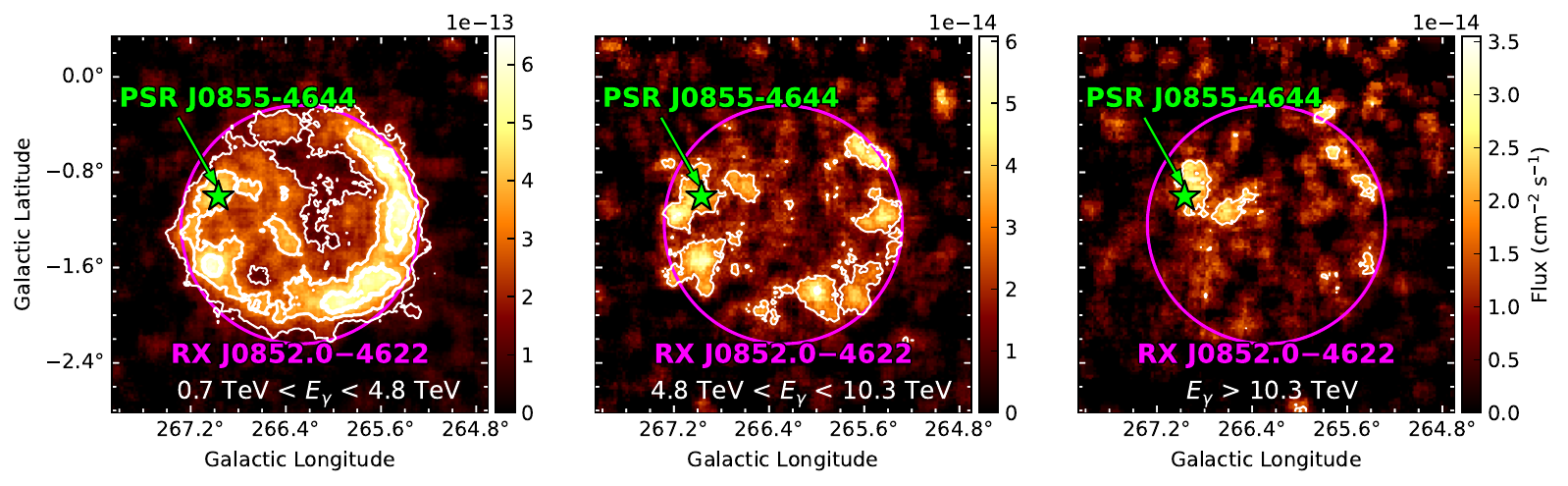}
\caption{TeV \gray flux map (cm$^{-2}$\,s$^{-1}$) of the Vela Junior region in three different energy bands. A smoothing radius of 0.1$^{\circ}$ is used. The Vela Junior SNR is indicated by the magenta circle, and \psr is indicated by the green star. The significance contours of the TeV $\gamma$-ray emission at the 5, 10, and 20$\sigma$ levels are shown in white, with increasing line width for increasing significance.}
\label{fig:edep_VJ}
\end{figure*}

Figure~\ref{fig:edep_VJ} shows energy-resolved \gray flux maps of the Vela Junior region in three energy bands: 0.7\,$<E_\gamma<$4.8\,TeV, 4.8\,$<E_\gamma<$10.3\,TeV, and $E_\gamma>$10.3\,TeV. The flux maps were generated assuming a power-law spectrum with an index of $-2$, and smoothed using a Gaussian kernel with a radius of 0.1$^{\circ}$. 
Both the extent of the SNR and the position of \psr are indicated. Vela Junior is indicated by the magenta circle, and the pulsar of interest in this study, \psr, is indicated by the green star. 
The flux maps show clear morphological changes in the emission as energy increases. In the lowest energy bin, the Vela Junior shell dominates, with the eastern rim appearing very bright. However, at energies above 10\,TeV, significant emission ($>5\sigma$) is observed coinciding with the position of \psr. 
This, along with a hard \gray region around the pulsar, suggests that the emission originates from \psr, possibly as a PWN (or halo). The \gray region is larger than the X-ray emission region, providing preliminary evidence that the electron population cools as particles propagate away from the pulsar.
This provides further motivation to study the region with a full spectro-morphological analysis.

\subsection{Spectro-morphological modelling}
\label{subsec:spectral} 
The TeV \gray emission in this region has a number of different structures (as showcased in Fig.~\ref{fig:edep_VJ}), which need to be modelled with care. 
While the well-defined SNR shell component (Vela Junior) has been established \citep{2018_hess_deep}, it is now possible to investigate the potential existence of additional sources using 3D analysis techniques by adding model components until no significant residual emission remains. 
First, we modelled the brightest objects in the region. Starting with the complex partial shell morphology at the eastern edge of Vela Junior, we utilised a template spatial model matching the morphology of TeV emission in this region (see Fig.~\ref{fig:A_template}). To build this model, we start with the flux map, as shown in the left panel of Fig.~\ref{fig:edep_VJ}. To enhance the most significant features and reduce any small-scale noise, the map was first smoothed using a Gaussian kernel with a width of 0.1$^{\circ}$. A threshold of $3.7\times10^{-13}$cm$^{-2}$s$^{-1}$ was applied, setting all values below this limit to zero, to remove any weaker emission and suppress low-intensity features. This step allowed us to model only the dominant emission features from this region, whilst minimising the influence of small fluctuations. The resulting map was normalised to a maximum value of one, then combined with an exponential cutoff power-law spectral model, hereafter referred to as component `A' (see Fig.~\ref{fig:A_template}).

\begin{figure}[hbt!]
\centering
\includegraphics[width=\columnwidth]{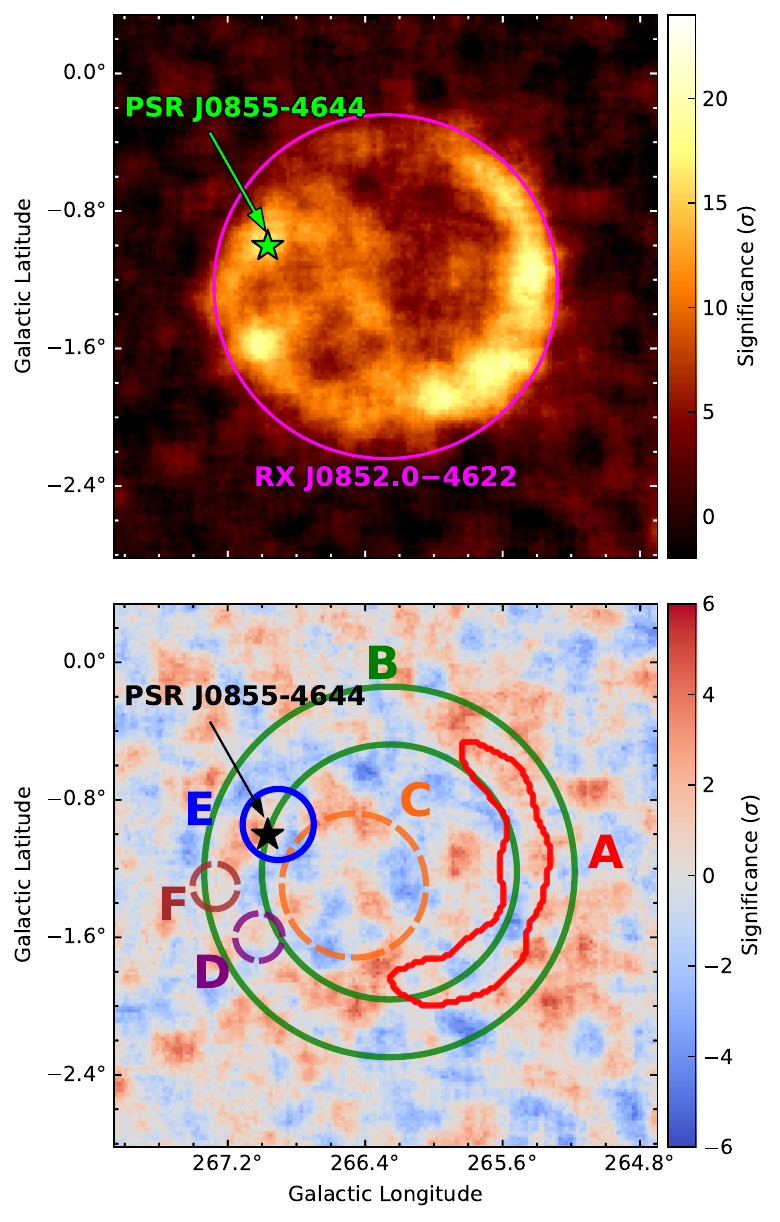}
\caption{Significance maps of \gray above 0.7\,TeV. \textit{Top:} Significance map with Vela Junior indicated by the magenta circle and \psr indicated by the green star. \textit{Bottom:} Residual significance map with the best-fit models overlaid. The six model components are shown in different colours, together with their respective labels. The circles represent the $\sim68$\% containment radii of the Gaussians.}
\label{fig:map_residuals}
\end{figure}

Next, a shell spatial model was used to describe the remaining contributions from the SNR shell. This model parametrises a uniformly emitting spherical shell projected in 2D. 
It is called component `B' and utilises the best-fit description from \cite{2018_hess_deep} as its initial values, with a shell radius of $1^{\circ}$ and shell thickness of $0.3^{\circ}$. 
Next, we iteratively added Gaussian components (named alphabetically from `C' to `F') with the initial position placed at the location of the highest significance peak. Each of these components were fitted using an exponential cutoff power-law spectral model. 
We stopped adding components when additional ones were no longer significant at the $4\sigma$ level above the background and the additional model was significantly offset from the source of interest within the FoV. This iterative addition of the models is shown in Fig.~\ref{fig:iterative_components}. 
Figure~\ref{fig:map_residuals} shows the \gray significance map (top) along with the residual significance map, after subtracting the best-fit models (bottom). We see no significant residual \gray emission ($>5\sigma$). 

For each component, we evaluated two nested hypotheses for the spectral model: 
\begin{itemize}
  \item $H_0$: Spectral parameters (index and cutoff) are shared with the shell model (component B, exponential cutoff power law), while normalisations are let free;
  \item $H_1$: Spectral parameters are independent, assuming a power-law.
\end{itemize}

We assessed these using the Wilks' theorem \citep{Wilks}. The spectra of the five components (`A' through `D' and `F') are compatible within statistics and are characterised by a power law with index $1.95\pm0.05_\textrm{stat}\pm0.18_\textrm{sys}$ and exponential cutoff energy $(10.25\pm0.94_\textrm{stat}\pm3.5_\textrm{sys}$)\,TeV; see Table~\ref{table:table_all}. It is important to note here that the spectrum of component B, corresponding to the SNR shell, is compatible with the spectrum published in \cite{2018_hess_deep}. The systematic uncertainties on our model parameters were derived through the methods outlined in Section\,2.1.3 of \cite{J1809_2023}. This Monte Carlo-based approach describes the systematics arising from discrepancies between the IRFs and the real observing conditions. In this approach, no effect is applied to the point spread function, so there is no contribution from the systematics associated with the position or extension. Instead, we quote the typical H.E.S.S. systematic pointing uncertainty of $20''$ per axis \citep{gillessen_07}. 

The comparison between $H_0$ and $H_1$ for component E yields a significance of $6.2\sigma$, indicating that its spectral parameters are not shared with the SNR and making it the only component that is clearly distinct from the others. Actually, the best-fit solution yields a cutoff parameter for component E that is statistically compatible with zero ($\lambda=0.025\pm0.02$\,TeV$^{-1}$). We therefore tested whether a spectral cutoff is statistically preferred. Comparing a power-law spectral model to an exponential cutoff power law, we find no significant preference ($\sim1\sigma$) for a cutoff in component E. 
The spectrum is therefore well described by a simple power law with a hard spectral index of $\Gamma_E = 1.81\pm0.07_\textrm{stat}$. 
The best-fit position of this component is $l_E=266.91^{\circ}\pm0.03_\textrm{stat}^{\circ}$ and $b_E=-0.94^{\circ}\pm0.02_\textrm{stat}^{\circ}$, with an extension of $\sigma_E=0.14^{\circ}\pm0.02_\textrm{stat}^{\circ}$. 
Component E has a detection significance of $12.2\sigma$. The extended source model is also significantly preferred over a point-like model (by $10.8\sigma$). 

\begin{figure}
\centering
\includegraphics[width=1\columnwidth]{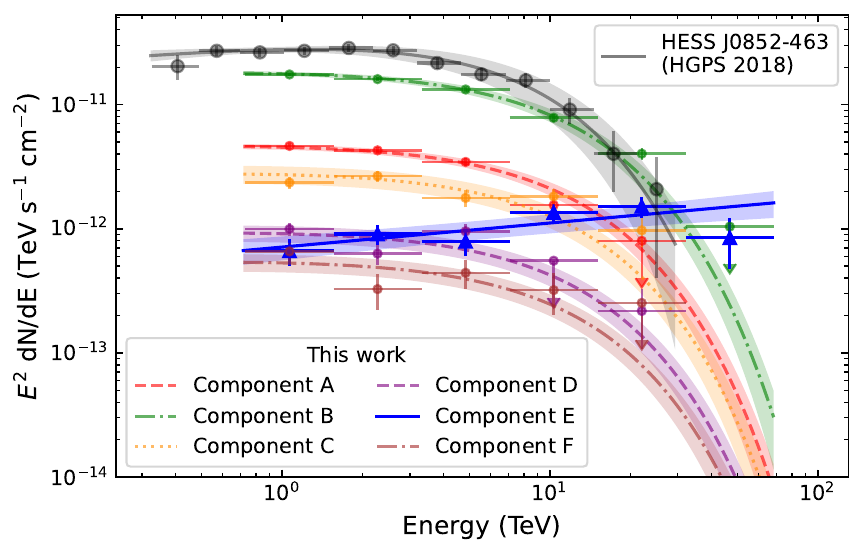}
\caption{Spectral energy distribution for each \hess component. The components from this work (`A' to `F') are shown in red (dashed), green (dot-dashed), orange (dotted), purple (dashed), blue (solid), and maroon (dot-dashed), respectively. The flux points are derived from the respective models. These spectra are compared to the \hessj result in solid black from \cite{hgps_2018}.}
\label{fig:spectra}
\end{figure}

\psr lies within the spatial extent of component E and therefore suggests a plausible \gray PWN. The offset between the pulsar position and emission centre is typical of a PWN, reflecting a non-homogeneous environment. 
Despite the absence of dedicated observation campaigns for Vela Junior since \cite{2018_hess_deep}, substantial improvements in the high-level analysis technique allowed us to obtain, for the first time, a clear source separation between the PWN of \psr and the SNR of Vela Junior. 
The spectra from each component, along with the \hessj spectrum from the HGPS \citep{hgps_2018}, are shown in Fig.~\ref{fig:spectra}, illustrating the different spectral shape of component E. The best-fit parameters for each model are provided in Table~\ref{table:table_all} and Table~\ref{table:table_E}.

\begin{table}
\centering
\caption{Best-fit model parameters for component E, modelled with a Gaussian spatial model and a power-law spectrum at a reference energy of 1\,TeV.}
\renewcommand{\arraystretch}{1.2} 
\begin{tabular}{lc}
\toprule[1.2pt]
Parameter & Value \\
\midrule
Galactic longitude [deg] & $266.91\pm0.03_\textrm{stat}\pm0.006_\textrm{sys}$ \\[0.3em]

Galactic latitude [deg] 
    & $-0.94\pm0.02_\textrm{stat}\pm0.006_\textrm{sys}$ \\

Extension [deg] 
    & $0.14\pm0.02_\textrm{stat}$ \\

Spectral index 
    & $1.81\pm0.07_\textrm{stat}\pm0.05_\textrm{sys}$ \\
    
\multirow{2}{*}{\parbox{10em}{Spectral flux (1\,TeV)\\{[$10^{-13}$ TeV$^{-1}$\,cm$^{-2}$\,s$^{-1}$]}}}
    & \multirow{2}{*}{\centering $(7.11\pm1.55_\textrm{stat}\pm0.75_\textrm{sys})$} \\ 
    & \\ 
    
\bottomrule[1.2pt]
\end{tabular}
\label{table:table_E}
\tablefoot{We quote the systematic uncertainties utilising the methods outlined in \cite{J1809_2023}, along with the $20''$ H.E.S.S. systematic pointing uncertainty \citep{gillessen_07}.}
\end{table}

\section{Discussion}
\label{discussion}
The following discussion focuses on component E, which is plausibly associated with the PWN of \psr. 
The parameters of all other components, utilised to subtract the Vela Junior emission and isolate the properties of the PWN, are not discussed further. 

\subsection{Emission characteristics from multi-wavelength observations}
\begin{figure}
\centering
\includegraphics[width=\columnwidth]{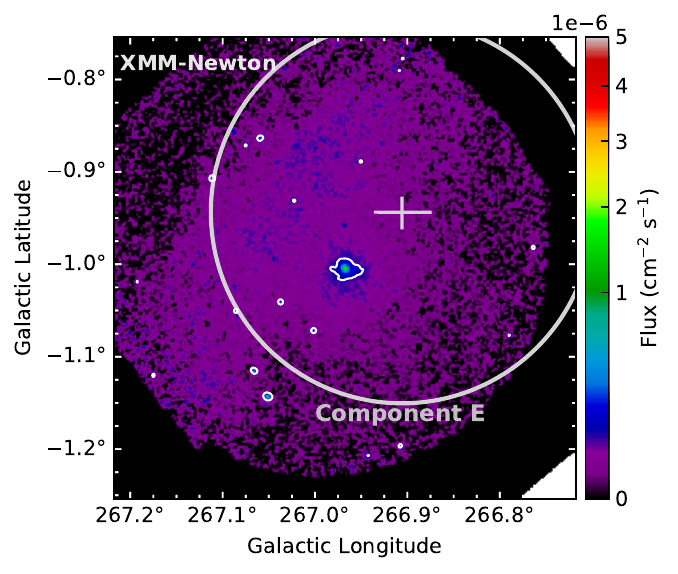}
\caption{XMM-Newton flux image from the MOS+PN camera in the high-energy band from 2-8\,keV. The map is smoothed with a Gaussian width of $4\arcsec$. The contour at $1.1\times10^{-7}$cm$^{-2}$s$^{-1}$ indicates the inner nebula of the PWN. The grey plus and circle indicate the position and size of component E in \grays, respectively. The diffuse emission to the north of the PWN is the rim of the Vela Junior remnant. The image is adapted from \cite{2013_acero_xrays}.}
\label{fig:XMM_map}
\end{figure}

The spatial positioning of component E strongly suggests that the TeV \gray emission is produced through the acceleration of particles from \psr as a PWN. A PWN is already detected at X-ray wavelengths with XMM-Newton \citep{2013_acero_xrays} in addition to a radio nebula \citep{2018_Maitra_radio}, both at the same location as component E (as shown in Fig.~\ref{fig:XMM_map}). 
The X-ray emission is non-thermal, with a spectral index that softens away from the pulsar (from $\Gamma=1.70{^{+0.07}_{-0.06}}$ to $\Gamma=2.01{\pm0.04}$), typical of an active PWN.
The measured TeV extension of $\sigma_E\sim (0.14^{\circ} \pm0.02_\textrm{stat}^{\circ})$ is an order of magnitude larger than the X-ray PWN, with a Gaussian width of $45''$ \citep[$\sim0.012^{\circ}$,][]{2013_acero_xrays}, as expected due to electron cooling timescales and transport effects. 
Figure~\ref{fig:XMM_map} shows the X-ray flux map from XMM-Newton in a high-energy band from 2-8\,keV, indicating the extent of the X-ray PWN and component E.
Given that PWNe are known \gray emitters \citep{hgps_2018,pwn_population_2018}, it is plausible to expect \gray emission from this source. 

Given that \psr also lies within component C, it is reasonable to question whether component C is also associated with \psr -- as an extended PWN or pulsar halo -- rather than with the SNR. We performed a number of tests, including fitting non-point-like models with varying sizes and position angles, to probe different morphological shapes for components E and C, both individually and jointly. 
We find no compelling evidence that component C constitutes extended emission from the PWN. 
The spectrum for component C is also statistically better associated with the SNR, as an exponential cutoff power-law spectral model (with fixed parameters from the shell model) is preferred over a power-law spectral model at a significance level of $6.4\sigma$. This indicates that the emission likely originates from the SNR rather than from the PWN. No significant residuals are observed at \gray energies (Fig.~\ref{fig:map_residuals}), implying that a potential PWN `tail' is not detected. 

The flux of component E at 1\,TeV is measured to be $7.11 \times 10^{-13}$\,TeV$^{-1}$\,cm$^{-2}$\,s$^{-1}$, which corresponds to a TeV luminosity between 1 and 10 TeV of $L_{\text{1-10\,TeV}} = 3.9 \times 10^{32}(d/\text{kpc})^2$\,erg\,s$^{-1}$.
Assuming a maximum distance of 900\,pc to the pulsar \citep{2013_acero_xrays}, this gives a TeV luminosity of $L_{\text{1-10\,TeV}} = 3.2 \times 10^{32}$\,erg\,s$^{-1}$.
The associated pulsar, \psr, has a spin-down power of $1.1 \times 10^{36}$\,erg\,s$^{-1}$, which leads to a TeV efficiency of 0.03\%, within the expected range for evolved PWNe \citep{pwn_population_2018}. 
The PWN interpretation is a natural explanation for the observed characteristics. 

The spectral index of component E is hard ($\Gamma_E \sim 1.8$), which is consistent with expectations for an energetic PWN. 
Hard \gray spectra at the cores of PWNe are also observed in other systems such as HESS\,J1825$-$137 \cite{J1825_Alison_2019}, where the innermost region of the nebula, centred on PSR\,B1823$-$13, has a hard spectral index of $\Gamma\lesssim2$. Vela\,X provides another example with a hard spectral index of $\Gamma=1.75$ close to the position of the pulsar \citep[0.3\,pc away,][]{VelaX_2019}. 

In 2011, Fermi-LAT (Fermi Large Area Telescope) reported the detection of the Vela Junior SNR; however, no component attached to the PWN has been detected \citep{fermi_2011}. Given the hard spectral index of the PWN component at TeV energies, this counterpart is unlikely to be detectable in the Fermi-LAT energy range. 
The emission from this component would be too faint at 1-10\,GeV due to the steepness of the spectral index. 

\subsection{Broadband keV to TeV joint fitting}
To understand the parent particle population better, we performed a joint multi-wavelength fit to both the X-ray and \gray emission using Naima \citep{naima} within the Gammapy framework, together with the Gammapy wrapper for the Sherpa X-ray modelling package \citep{sherpa} and the XSPEC spectral models \citep{xspec}.
For the X-ray data, we used observations from the XMM-Newton PN camera, as detailed in \cite{2013_acero_xrays}. 
A mask was applied to this dataset to restrict the fit only to the emission from high-energy electrons, focusing on the 2-8\,keV range (consistent with Fig.~\ref{fig:XMM_map}) to avoid thermal contamination from the Vela SNR and the rim of Vela Junior.

The emission is described by a single-zone model, in which the population of particles follows a power-law spectral model with an exponential cutoff, intended as a phenomenological description.
The synchrotron emission was modelled based on the ambient magnetic field strength, $B$, the parent particle population, and a component accounting for photoelectric absorption. The absorption column density was taken to be $N_H=7.6\times10^{21}$cm$^{-2}$ as per \cite{2013_acero_xrays} and fixed during the fit. 
The same electron population was assumed to be responsible for the \gray emission resulting from IC scattering of the cosmic microwave background (CMB), 
near-infrared (NIR), and far-infrared (FIR) background photon fields. 
For the FIR component, we assumed a temperature of 30\,K with an energy density of 0.5\,eV\,cm$^{-3}$, while for the NIR component we assumed a temperature of 3000\,K with an energy density of 1\,eV\,cm$^{-3}$. These values are typical of the interstellar radiation field conditions outside the inner Galactic region \citep[see for example,][]{popescu_2017,porter_2017}. The distance to the source was assumed to be 900\,pc from X-ray observations \citep{2013_acero_xrays}. 

\begin{figure}
\centering
\includegraphics[width=1\columnwidth]{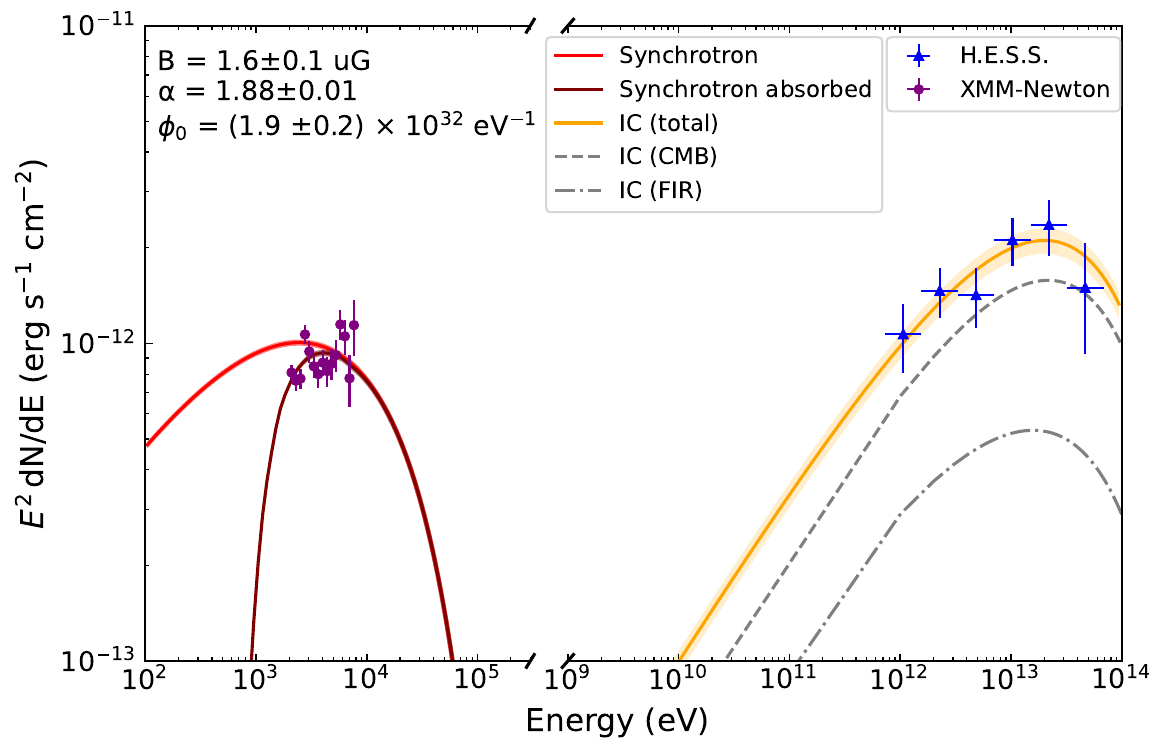}
\caption{Broadband spectral energy distribution of component E in a simple one-zone leptonic scenario. The IC emission is shown for the CMB (grey-dashed line) and FIR (dot-dashed line) components. The total IC emission is indicated by the orange line. The NIR cannot be seen on this plot as it is a subdominant component.
The red and maroon lines represent the synchrotron emission and synchrotron emission with absorption, respectively.
An X-ray absorption column density of $N_H=7.6\times10^{21}$cm$^{-2}$ is assumed for this model. The XMM-Newton X-ray flux points are shown by the dark-purple-filled squares, and the TeV \gray flux points from component E are indicated by the blue-filled triangles. Both sets of flux points were calculated from the one-zone single model.}
\label{fig:sed_wide}
\end{figure}

The power-law spectral model of the \hess component E from Sect.~\ref{subsec:spectral} was replaced by the IC component. The synchrotron component was utilised to describe the X-ray emission, extracted from a circular region with a radius of 0.14$^{\circ}$ representing component E. The electron spectral index, amplitude, and magnetic field were fitted simultaneously across the X-ray (XMM-Newton) and \gray (\hess) data. This joint fit was performed on the 1D spectral dataset obtained from XMM-Newton and the 3D dataset from \hess, enabling the use of the forward-folding technique.
The results of this fit are shown in Table~\ref{table:table_sed} and illustrated in Fig.~\ref{fig:sed_wide}. 
The flux points shown in Fig.~\ref{fig:sed_wide} were then derived assuming, in the individual energy bins, the spectral model derived from the fit.

\begin{table}[htb]
\centering
\caption{Best-fit model parameters for a simple one-zone model describing the emission from both X-rays and \grays.}
\begin{tabular}{ccc}
\toprule[1.2pt]
Parameter & Description & Value \\
\midrule
 $B$ &  Magnetic field & $1.6\pm0.1\mu$G \\
 $\alpha$ & Index & $1.88\pm0.04$ \\
 $\phi_0$ & Normalisation & $(1.9\pm0.2) \times 10^{32}$eV$^{-1}$ \\
\bottomrule[1.2pt]
\end{tabular}
\label{table:table_sed}
\tablefoot{The column density is frozen at $N_H=7.6\times10^{21}$cm$^{-2}$.
The magnetic field, $B$, is taken to be the lower limit; the spectral index and normalisation of the parent particle population are $\alpha$ and $\phi_0$, respectively.}
\end{table}

The model of the broadband spectral energy distribution in Fig.~\ref{fig:sed_wide} describes the H.E.S.S. and XMM-Newton data well. 
We find an electron spectral index of $\alpha\sim1.9$, obtained from the joint fit, which is a reasonable value for the injection spectrum (PWNe are expected to exhibit a hard spectral indices; see \cite{jager_2008,olmi_2023}). 
The magnetic field of $B \sim 1.6\mu$G represents an effective average value over the large region of interest. Although this value is relatively low, it is not uncommon when fitting the X-ray and TeV data with a simple stationary one-zone model \citep[see][]{j1303_Bfield_2012}.
As shown in previous studies \citep{gelfand_2009, vanetten_romani_2011, collins_2024}, the magnetic field in a PWN decreases with distance beyond the terminal shock, resulting in a low average magnetic field strength over the integration region. In addition, continuously accelerated particles suffer from radiative and adiabatic losses, such that most of the highly energetic particles that emit X-rays lie closer to the termination shock than those producing emission beyond 1\,TeV \citep[e.g.][]{J1809_2023}.

\section{Conclusions}
\label{conclusions}
The X-ray bright PWN powered by \psr overlaps with the Vela Junior shell, making this a complex analysis region. 
Using this novel 3D analysis technique with an accurate description of the spectro-morphological parameters allowed us to disentangle the two sources for the first time. We find a \gray PWN component (component E), positionally consistent with the X-ray PWN. This extended emission ($\sigma_E=0.14^{\circ}\pm0.02_\textrm{stat}^{\circ}$) is detected at a significance level of $12.2\sigma$.

A detailed spectral analysis revealed that component E exhibits different spectral characteristics than the other components, which are likely associated with the SNR itself. Component E is described by a power-law spectral model with a hard index of $\Gamma_E = 1.81\pm0.07_\textrm{stat}$. This spectral difference suggests that this component does not originate from the SNR but rather from the pulsar within its vicinity. 
Given the observed characteristics, spectral properties, and the spatial alignment with the X-ray PWN, interpreting this component as a PWN is the most plausible explanation. 
Using the spin-down power ($\dot{E}=1.1\times10^{36}$\,erg\,s$^{-1}$) along with the distance to the pulsar (900\,pc) allows us to deduce \mbox{$\dot{E}/D^2 > 10^{34}\,\mathrm{erg\,s^{-1}\,kpc^{-2}}$}, which is characteristic of the pulsars known to power PWNe at TeV energies \citep{pwn_population_2018}.
This scenario is also supported from an energetics point of view by the TeV conversion efficiency of the PWN, 0.03\%. Given the other characteristics of the pulsar, this PWN is also well placed within the population of TeV PWNe presented by \cite{pwn_population_2018}.

A leptonic scenario was considered, in which a synchrotron component and an IC component were used to describe the emission at X-ray and \gray wavelengths, respectively. A joint fit of the XMM-Newton X-ray data and the \hess \gray data reveals a spectral index of $\alpha=1.88\pm0.01$ for the electron population and allowed us to place a lower limit on the magnetic field strength of $1.6\mu$G. 

\section{Data availability}
The scientific data products presented in this paper are publicly accessible via Zenodo \href{https://zenodo.org/records/18351700}{doi:10.5281/zenodo.18351700} \citep{data_zenodo}.

\begin{acknowledgements}
The support of the Namibian authorities and of the University of Namibia in facilitating the construction and operation of H.E.S.S. is gratefully acknowledged, as is the support by the German Ministry for Education and Research (BMBF), the Max Planck Society, the Helmholtz Association, the French Ministry of Higher Education, Research and Innovation, the Centre National de la Recherche Scientifique (CNRS/IN2P3 and CNRS/INSU), the Commissariat à l’énergie atomique et aux énergies alternatives (CEA), the U.K. Science and Technology Facilities Council (STFC), the Polish Ministry of Education and Science, agreement no. 2021/WK/06, the South African Department of Science and Innovation and National Research Foundation, the University of Namibia, the National Commission on Research, Science \& Technology of Namibia (NCRST), the Austrian Federal Ministry of Education, Science and Research and the Austrian Science Fund (FWF), the Australian Research Council (ARC), the Japan Society for the Promotion of Science, the University of Amsterdam and the Science Committee of Armenia grant 21AG-1C085. We appreciate the excellent work of the technical support staff in Berlin, Zeuthen, Heidelberg, Palaiseau, Paris, Saclay, Tübingen and in Namibia in the construction and operation of the equipment. This work benefited from services provided by the H.E.S.S. Virtual Organisation, supported by the national resource
providers of the EGI Federation.
\end{acknowledgements}

\bibliographystyle{aa}
\bibliography{bibliography}

\begin{appendix} 

\section{3D modelling details}
\label{3D-details}

\begin{figure}[hbt!]
\centering
\includegraphics[width=\columnwidth]{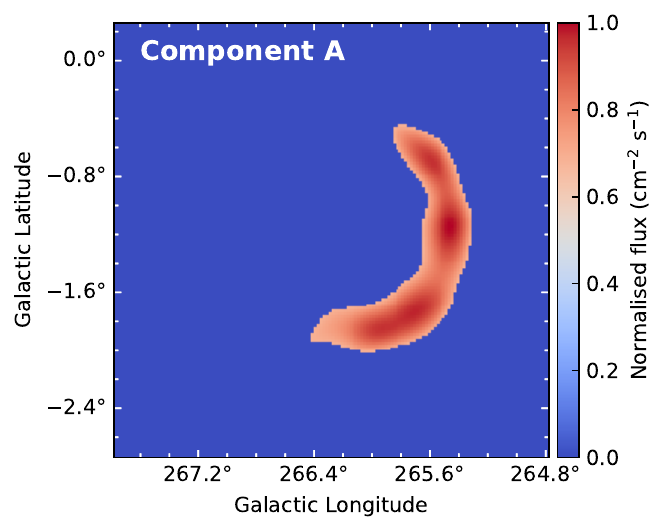}
\caption{Template for component A: Representation of the eastern shell.}
\label{fig:A_template}
\end{figure}

Figure~\ref{fig:A_template} shows the template spatial model utilised to represent the bright emission from the eastern edge of the Vela Junior SNR. This is called component A throughout the text. 

Table~\ref{table:table_all} shows the results of the best-fit models for each of the components. 

\begin{table*}[!b]
\centering
\caption{Best-fit model parameters for components A through F.}
\begin{tabular}{ccccc}
\toprule[1.2pt]
Component & Galactic longitude & Galactic latitude & Extension & Spectral flux (1\,TeV) \\

 & [deg] & [deg] & [deg] & [$10^{-12}$ TeV$^{-1}$\,cm$^{-2}$\,s$^{-1}$] \\
\midrule
A & - & - & - & $
(5.0\pm0.2_\textrm{stat}\pm0.4_\textrm{sys})$ \\ 

B & $266.25\pm0.006_\textrm{sys}$ & $-1.22\pm0.006_\textrm{sys}$ & 1.067 & 
$(19.5\pm0.7_\textrm{stat}\pm1.4_\textrm{sys})$\\ 

C & $266.47\pm0.03_\textrm{stat}\pm0.006_\textrm{sys}$ & $-1.30\pm0.04_\textrm{stat}\pm0.006_\textrm{sys}$ & $0.28\pm0.03_\textrm{stat}$ & 
$(3.0\pm0.5_\textrm{stat}\pm 0.2_\textrm{sys})$\\ 

D & $267.02\pm0.01_\textrm{stat}\pm0.006_\textrm{sys}$ & $-1.59\pm0.01_\textrm{stat}\pm0.006_\textrm{sys}$ & $0.09\pm0.01_\textrm{stat}$ & 
$(1.0 \pm0.15_\textrm{stat}\pm0.08_\textrm{sys})$ \\ 

F & $267.27\pm0.01_\textrm{stat}\pm0.006_\textrm{sys}$ & $-1.31\pm0.02_\textrm{stat}\pm0.006_\textrm{sys}$ & $0.09\pm0.01_\textrm{stat}$ & 
$(0.584 \pm0.094_\textrm{stat}\pm0.045_\textrm{sys})$ \\ 
\bottomrule[1.2pt]
\end{tabular}
\label{table:table_all}
\tablefoot{These components are described by a spectral index of $1.95\pm0.03_\textrm{stat}\pm0.18_\textrm{sys}$ and an exponential cutoff energy of $(10.25\pm0.94_\textrm{stat}\pm3.5_\textrm{sys}$)\,TeV at a reference energy of 1\,TeV. We quote the systematic uncertainties utilising the methods outlined in \cite{J1809_2023}, along with the $20''$ systematic pointing uncertainty of H.E.S.S..}
\end{table*}

\begin{figure*}
\centering
\includegraphics[width=1.1\columnwidth]{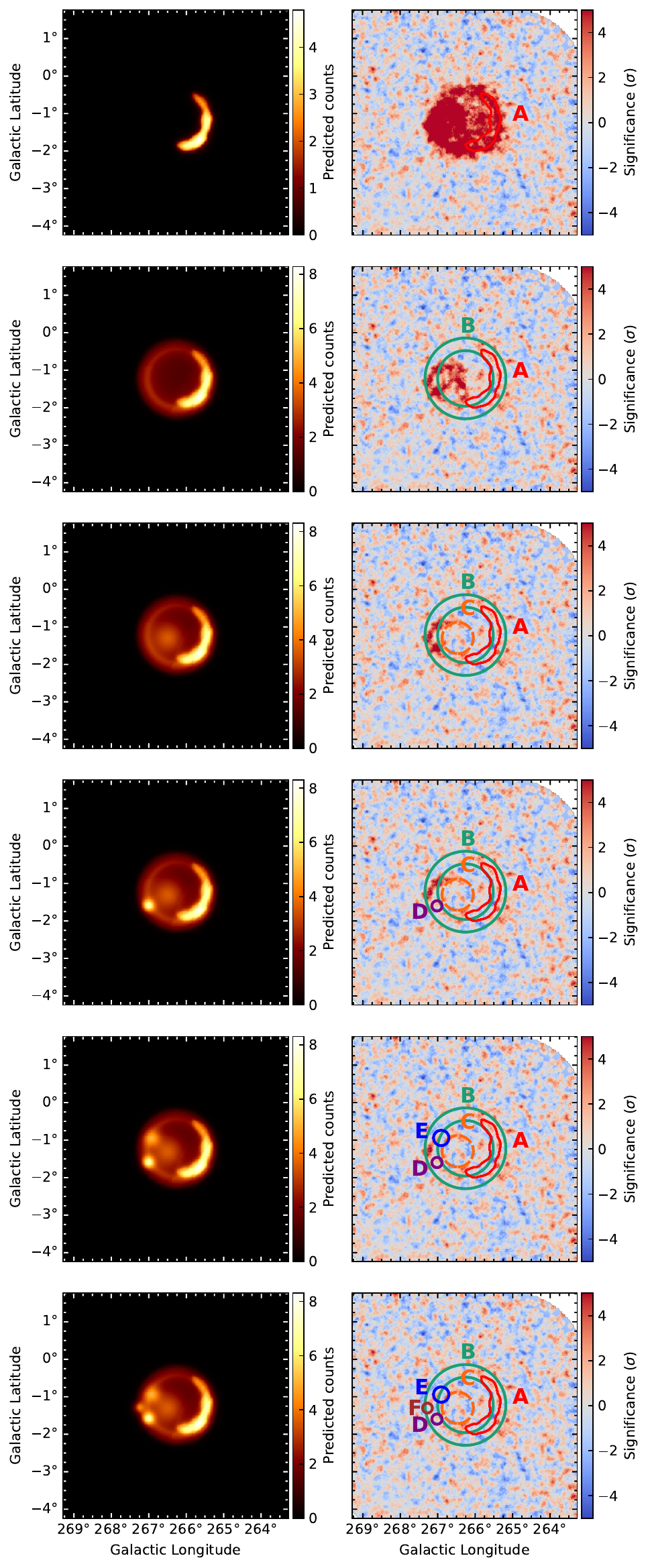}
\caption{Iterative addition of the models outlined in Sect.~\ref{subsec:spectral} from `A' through to `F' in red, green, orange, purple, blue, and maroon, respectively. \textit{Left}: Predicted counts for the models. \textit{Right}: Residual significance maps.}
\label{fig:iterative_components}
\end{figure*}

Figure~\ref{fig:iterative_components} shows the iterative addition of the models described in Sect.~\ref{subsec:spectral}.

\end{appendix}

\end{document}